\documentclass[aps,prx,twocolumn,floatfix,amsmath,amssymb]{revtex4-2}

\usepackage{float}
\usepackage{graphicx}

\begin{document}

\title{Geometric Parameterization of Kraus Operators with Applications to Quasi Inverse Channels for Multi Qubit Systems}%

\author{Zain Ateeq}
\affiliation{Department of Physics, Lahore University of Management Sciences, Lahore 54792, Pakistan}

\author{Muhammad Faryad}
\email{muhammad.faryad@lums.edu.pk}
\affiliation{Department of Physics, Lahore University of Management Sciences, Lahore 54792, Pakistan}

\date{August 2025}

\begin{abstract}

This work presents a differentiable geometric parameterization of quantum channels in Kraus representation, which can be efficiently probed to find an unknown quantum channel. We explore its feasibility in finding the quasi inverse channels, which can be a tedious analytically for complex noise processes and is often achievable only for a limited range of parameters. In this regard, machine learning based algorithms have been employed successfully to find quasi inverse of quantum channels. The space of quantum channels in this scheme is a unit hypersphere, and components of mutually constrained unit vectors residing in this space, are used to construct a physically valid quantum channel. Symplectic constraints, orthogonality, and unit length of the vectors suffice to maintain complete positivity and the trace-preserving property of the channels. By performing gradient descent on this parametric space with a fidelity-based loss function, this approach is found to optimize quasi inverse of a variety of quantum channels, not limited to single-qubits, proving its effectiveness. 
\end{abstract}

\maketitle
\tableofcontents

\section{Introduction}
The dynamics of open quantum systems, in the most general fashion, is described by completely positive trace preserving (CPTP) maps \cite{nielsen}, also called quantum channels. Quantum channels are crucial for understanding the interaction between a system and its environment as they characterize unitary evolution, decoherence, noise, and measurements \cite{breuer}. Understanding and characterizing unknown quantum channels, as well as investigating the quasi inverse of known or unknown quantum channels, is an active area of research. However, obtaining analytical expressions for quasi inverse channels is limited to specific cases and often becomes intractable for higher dimensions \cite{kretschmann}. This is what motivates machine learning–based approaches which provide a promising alternative, offering flexible and data-driven frameworks to obtain approximate quasi inverses and generalizing beyond analytically solvable regimes \cite{carrasquilla}\cite{aziz2025}.

Several parameterization strategies exist for modeling CPTP quantum channels, each with their advantages and limitations. Traditional approaches, such as Cholesky decomposition of the Choi matrix, ensure complete positivity, but is not trace preserving inherently \cite{choi}, limiting automation and flexibility. Stinespring dilation offers a physically intuitive framework that naturally preserves the CPTP structure \cite{stinespring}, but introduces a high computational overhead due to the enlarged Hilbert space. Orthonormal basis expansions (for example, Pauli or Gell-Mann bases) are analytically tractable and useful for theoretical insights, although they rely on loosely enforced CPTP constraints \cite{nielsen}. In contrast, gradient projection techniques, while simple to implement, suffer from instability and slow convergence when applied to constrained optimization \cite{bertsekas}. In this work, we present an approach based on differentiable parameterization using vectors in hyperspherical coordinates, which maintains the CPTP structure by construction. It offers a compact and geometrically motivated framework suitable for gradient-based learning. This geometric structure with inherent CPTP preservation and differentiability distinguishes it from existing parameterization schemes.

This parameterization scheme for CPTP maps provides a Bloch sphere analogue for Kraus operators of the Hilbert space of arbitrary dimension $d$. Any CPTP map can be expressed in terms of a set of Kraus operators \( \{K_i\} \), satisfying the completeness condition $\sum_i (K^i)^\dagger (K^i) = I$ \cite{nielsen}. In this picture, every possible Kraus operator lives in a hypersphere, which we call the Kraus sphere. In this space, components of $d$ constrained vectors, we call them Kraus vectors, give us a set of CPTP Kraus operators. The constraints on these vectors come from the completeness relation. The fact that Kraus vectors are unit vectors and two other constraints, namely euclidean orthogonality and symplectic orthogonality on Kraus vectors, ensure the maps to be trace preserving and completely positive. In this way, we can span the whole space of Kraus operators via probing the Kraus sphere using Kraus vectors. To accomplish this, we derived the Lie algebra of the group of transformations for Kraus vectors that preserve the orthogonality and symplectic constraints. From these generators, we work out the generalized finite transformation matrices, parameterized by angles, to transform Kraus vectors of any $n$-qubit system.

The proposed parameterization of Kraus operators gives a flexible framework for exploring the space of CPTP quantum channels. With an in-built differentiable and constraint-preserving structure, this approach enables efficient navigation of the quantum channel manifold using the gradient based optimization. Theoretically, it facilitates the study of the geometric and algebraic structure of CPTP maps and can possibly give insight in the analysis of extremal channels, convex combinations, and quasi inverses, which are relevant in the context of approximate quantum error recovery. Numerically, this formulation is particularly well suited in Monte Carlo sampling on CPTP maps, integration into variational algorithms, machine learning models, and channel learning tasks, where differentiability is essential for backpropagation and optimization.

Using this parameterization, we formulate an approach to learn quasi inverse of quantum channels as a fidelity maximization task. Given a set of original states sampled randomly from the Bloch ball, along with a set of transformed states resulting from passing the original set from an unknown quantum channel, our goal is to recover the original states as closely as possible by inverting the transformed states. The approximate inverse channel, which maps transformed states to recovered states, is the quasi inverse of the unknown channel. In the end, we want to achieve the maximum possible fidelity between the original and recovered states. We get to this quasi inverse by performing a gradient decent on the Kraus sphere with a fidelity-based loss function, and the parameterization ensures that the optimized quasi inverse is always a CPTP map. We demonstrate that this method converges to high-fidelity recovering quasi inverses for general quantum channels.

The following is the plan of the paper: Section \ref{sec:parameterization} explains how we parameterize the Kraus operators. We discuss the case of a single-qubit system and explain how the Kraus vectors are held on the Kraus sphere with normalization, orthogonality, and symplectic constraints. Furthermore, we describe how to generalize this approach to $n$-qubits and produce a random set of Kraus operators for any dimensional system. Section \ref{sec:generators_transformations} explains how we use the orthogonal and symplectic symetries of Kraus vectors to develop the Lie algebra behind the transformations, which preserves the CPTP structure of Kraus vectors and derive a general expression for $n$-qubit finite transformations which enables us to span the entire space of CPTP Channels. In Section \ref{sec:methodology}, we introduce our approach to learn the quasi inverse using the gradient decent. We explain the loss function along with the method we employed to compute its gradient on the Kraus sphere. Section \ref{sec:results} presents the results we obtained for one- and two-qubit systems using the fidelity comparison before and after applying the quasi inverse. In the end, Section \ref{sec:conclusions} discusses the conclusions and the future prospects of the work.

\section{\label{sec:parameterization}Parameterization of Kraus Operators }
The Kraus operators of an $n$-qubit channel live in a Hilbert space of size $d=2^n$ and require a $d^2$ number of Kraus operators $K^{\alpha}$, each being a $d \times d$ complex matrix \cite{nielsen}. This would make a total of $d^2\times d \times d = d^4$ complex parameters, or equivalently $2d^4$ real parameters. However, choosing values for these real parameters has restrictions and not every choice yields a physically valid quantum channel. In order for the channel to be CPTP, the Kraus operators must satisfy the \textit{completeness relation} \cite{nielsen}:
\begin{eqnarray}
    \sum_{\alpha=1}^{d^2} (K^{\alpha})^\dagger (K^{\alpha}) = I.
    \label{eq:pko1}
\end{eqnarray}
Entries of $K^{\alpha}$ are complex numbers which can be written as
\begin{eqnarray}
    (K^{\alpha})_{ij} = x^{\alpha}_{ij} + i y^{\alpha}_{ij}.
    \label{eq:pko3}
\end{eqnarray}
By inserting this into Eq. \ref{eq:pko1} and solving for the expression of a single matrix entry we get
\begin{eqnarray}
\sum\limits_{\alpha=1}^{d^2}\sum\limits_{k=1}^d\left[x^\alpha_{ki}x^\alpha_{kj}+y^\alpha_{ki}y^\alpha_{kj}+i(x^\alpha_{ki}y^\alpha_{kj}-y^\alpha_{ki}x^\alpha_{kj})\right]=\delta_{ij},
\label{eq:pkom}
\end{eqnarray}
where $\delta_{ij}$ is the Kronecker delta. This equation contains all the constraints required to make a set of Kraus operators a valid CPTP map. 

\subsection{Single-Qubit System}
For a single-qubit system we have $d=2$, thus four $2 \times 2$ kraus operators. 
For the diagonal entries when $i=j$ we get two normalization constraints from Eq. (\ref{eq:pkom}) as follows:
\begin{eqnarray}
    \sum\limits_{\alpha=1}^4\sum\limits_{k=1}^2 \left[(x_{k1}^{\alpha})^2 + (y_{k1}^{\alpha})^2\right] = 1,
    \label{eq:pko5}
\end{eqnarray}
and
\begin{eqnarray}
\sum\limits_{\alpha=1}^4\sum\limits_{k=1}^2 \left[(x_{k2}^{\alpha})^2 + (y_{k2}^{\alpha})^2\right] = 1.
\label{eq:pko6}
\end{eqnarray}
For $i \neq j$, we get,
\begin{eqnarray}
\sum\limits_{\alpha=1}^4\sum\limits_{k=1}^2 \big[x^{\alpha}_{k1} x^{\alpha}_{k2} + y^{\alpha}_{k1} y^{\alpha}_{k2} + 
i (x^{\alpha}_{k1}y^{\alpha}_{k2} - y^{\alpha}_{k1}x^{\alpha}_{k2})\big]=0.
\label{eq:pko7}
\end{eqnarray}
From Eq. (\ref{eq:pko7}), we can further write
\begin{eqnarray}
\sum\limits_{\alpha=1}^4\sum\limits_{k=1}^2 (x^{\alpha}_{k1} x^{\alpha}_{k2} + y^{\alpha}_{k1} y^{\alpha}_{k2}) = 0,
\label{eq:pko9}
\end{eqnarray}
\begin{eqnarray}
\sum\limits_{\alpha=1}^4\sum\limits_{k=1}^2(x^{\alpha}_{k1}y^{\alpha}_{k2} - y^{\alpha}_{k1}x^{\alpha}_{k2}) = 0.
\label{eq:pko10}
\end{eqnarray}

The structure of Eq. (\ref{eq:pko5}) and Eq. (\ref{eq:pko6}) naturally allows their interpretation as the squared norms of two distinct vectors, say \( \mathbf{v} \) and \( \mathbf{w} \), residing in \( \mathbb{R}^{16} \). Now Eq. (\ref{eq:pko5}) and Eq. (\ref{eq:pko6}) reduce to normalization constraints \( \|\mathbf{v}\|^2 = 1 \) and \( \|\mathbf{w}\|^2 = 1 \) respectively. Hence defining:

\begin{eqnarray}
    \sum\limits_{\alpha=1}^4\sum\limits_{k=1}^2 \left[(x_{k1}^{\alpha})^2 + (y_{k1}^{\alpha})^2\right] =\sum\limits_{n=1}^{16}v_n^2=\|\mathbf{v}\|^2 =1,
    \label{eq:normv}
\end{eqnarray}
\begin{eqnarray}
\sum\limits_{\alpha=1}^4\sum\limits_{k=1}^2 \left[(x_{k2}^{\alpha})^2 + (y_{k2}^{\alpha})^2\right] =\sum\limits_{n=1}^{16} w_n^2= \|\mathbf{w}\|^2=1,
\label{eq:normw}
\end{eqnarray}
where $v_n$ are the components of $\mathbf{v}$ and $w_n$ are the components of $\mathbf{w}$. We call these vectors the Kraus vectors. In this way, we get the following relabeling of variables:
\begin{eqnarray}
\mathbf{v} =
\left[
\begin{array}{c}
v_1 \\ v_2 \\ v_3 \\ v_4 \\
v_5 \\ v_6 \\ v_7 \\ v_8 \\
v_9 \\ v_{10} \\ v_{11} \\ v_{12} \\
v_{13} \\ v_{14} \\ v_{15} \\ v_{16}
\end{array}
\right]
=
\left[
\begin{array}{c}
x^1_{11} \\ y^1_{11} \\ x^1_{21} \\ y^1_{21} \\
x^2_{11} \\ y^2_{11} \\ x^2_{21} \\ y^2_{21} \\
x^3_{11} \\ y^3_{11} \\ x^3_{21} \\ y^3_{21} \\
x^4_{11} \\ y^4_{11} \\ x^4_{21} \\ y^4_{21}
\end{array}
\right],
\mathbf{w}=
\left[
\begin{array}{c}
w_1 \\ w_2 \\ w_3 \\ w_4 \\
w_5 \\ w_6 \\ w_7 \\ w_8 \\
w_9 \\ w_{10} \\ w_{11} \\ w_{12} \\
w_{13} \\ w_{14} \\ w_{15} \\ w_{16}
\end{array}
\right]
=
\left[
\begin{array}{c}
x^{1}_{12} \\ y^{1}_{12} \\ x^{1}_{22} \\ y^{1}_{22} \\
x^{2}_{12} \\ y^{2}_{12} \\ x^{2}_{22} \\ y^{2}_{22} \\
x^{3}_{12} \\ y^{3}_{12} \\ x^{3}_{22} \\ y^{3}_{22} \\
x^{4}_{12} \\ y^{4}_{12} \\ x^{4}_{22} \\ y^{4}_{22}
\end{array}
\right].
\label{eq:relabelW}\end{eqnarray}
After writing Eq. (\ref{eq:pko9}) and Eq. (\ref{eq:pko10}) in this labeling we get 
\begin{eqnarray}
\sum\limits_{\alpha=1}^4\sum\limits_{k=1}^2 (x^{\alpha}_{k1} x^{\alpha}_{k2} + y^{\alpha}_{k1} y^{\alpha}_{k2}) \nonumber \\
=v_1w_1+v_2w_2+&\cdots+v_{16}w_{16}=0,
\label{eq:pko14}
\end{eqnarray}
and
\begin{eqnarray}
\sum\limits_{\alpha=1}^4\sum\limits_{k=1}^2(x^{\alpha}_{k1}y^{\alpha}_{k2} - y^{\alpha}_{k1}x^{\alpha}_{k2}) \nonumber \\
= v_1w_2-v_2w_1+v_3w_4-&\cdots-v_{16}w_{15}=0.
\label{eq:pko15}
\end{eqnarray}
This reveals a rich structure as we can reduce right hand side of Eq. (\ref{eq:pko14}) to a standard Euclidean inner product of $\mathbf{v}$ and $\mathbf{w}$,
\begin{eqnarray}
\sum\limits_{\alpha=1}^4\sum\limits_{k=1}^2 (x^{\alpha}_{k1} x^{\alpha}_{k2} + y^{\alpha}_{k1} y^{\alpha}_{k2})=\mathbf{v}\cdot\mathbf{w}=0,
\label{eq:dotproduct}
\end{eqnarray}
indicating that the two vectors \( \mathbf{v} \) and \( \mathbf{w} \) are orthogonal in \( \mathbb{R}^{16} \). 

Now right hand side of Eq. (\ref{eq:pko15}), originally involving alternating cross terms can be written as
\begin{eqnarray}
\sum_{k=1}^{8} \left( v_{2k-1} w_{2k} - v_{2k} w_{2k-1} \right) = 0.
\label{eq:pko17}
\end{eqnarray}
This antisymmetric combination can be viewed as a skew inner product or a symplectic inner product between \( \mathbf{v} \) and \( \mathbf{w} \) with respect to a symplectic bilinear form $S$. The symplectic bilinear form is defined as a matrix \( S \in \mathbb{R}^{2n \times 2n} \) that satisfies the condition:
\begin{eqnarray}
S^\top = -S.
\label{eq:symplectic_condition}
\end{eqnarray}
This structure allows us to write Eq. (\ref{eq:pko17}) as
\begin{eqnarray}
\sum_{k=1}^{8} \left( v_{2k-1} w_{2k} - v_{2k} w_{2k-1} \right) =\mathbf{v}\cdot S\mathbf{w}= 0,
\label{eq:pko17symp}
\end{eqnarray}
where
\begin{eqnarray}
S = I_8 \otimes
\left[
\begin{array}{cc}
0 & 1 \\
-1 & 0
\end{array}
\right].
\label{eq:pko19}
\end{eqnarray}
Eq. (\ref{eq:pko7}) now takes the following form
\begin{eqnarray}
\mathbf{v} \cdot\mathbf{w} + i(\mathbf{v}\cdot S\mathbf{w})=0,
\label{eq:offdiag}
\end{eqnarray}
and now we can write the completeness relation for single-qubit systems by using Eq. (\ref{eq:normv}), (\ref{eq:normw}) and (\ref{eq:offdiag}) as
\begin{eqnarray}
\left[
\begin{array}{cc}
\|\mathbf{v}\|^2 & \mathbf{v} \cdot\mathbf{w} + i(\mathbf{v}\cdot S\mathbf{w}) \\
\mathbf{w} \cdot\mathbf{v} + i(\mathbf{w}\cdot S\mathbf{v}) & \|\mathbf{w}\|^2
\end{array}
\right]
=I.
\label{eq:pkoM}
\end{eqnarray}
This is the completeness relation translated in terms of Kraus vectors.

\subsection{Multi-Qubit System}
In the same way, we can generalize this approach to $n$-qubits. For $i=j$ we get $d$ number of diagonal terms which we write using vector nomrs as
\begin{eqnarray}
    \sum\limits_{\alpha=1}^{d^2}\sum\limits_{k=1}^d \left[(x_{ki}^{\alpha})^2 + (y_{ki}^{\alpha})^2\right]=\|\mathbf{v}_i\|^2 =1.
    \label{eq:generalnorm}
\end{eqnarray}
This dictates that for $n$-qubit system, we get $d$ number of Kraus vectors. 
For $i \neq j$ the terms define the euclidean and symplectic orthogonality among the Kraus vectors as
\begin{eqnarray}
    \sum\limits_{\alpha=1}^{d^2}\sum\limits_{k=1}^d \big[x^{\alpha}_{ki} x^{\alpha}_{kj} + y^{\alpha}_{ki} y^{\alpha}_{kj} + 
i (x^{\alpha}_{ki}y^{\alpha}_{kj} - y^{\alpha}_{ki}x^{\alpha}_{kj})\big] \nonumber\\
=\mathbf{v}_i\cdot\mathbf{v}_j+i(\mathbf{v}_i\cdot S\mathbf{v}_j)=0.
\label{eq:eucsymp}
\end{eqnarray}
By combining Eq. (\ref{eq:generalnorm}) and (\ref{eq:eucsymp}) we get
\begin{eqnarray}
\mathbf{v}_i\cdot\mathbf{v}_j+i(\mathbf{v}_i\cdot S\mathbf{v}_j)=\delta_{ij}
\label{eq:ncomprel}
\end{eqnarray}
This is the completeness relation for the $n$-qubit system in terms of Kraus vectors. The dimensionality of the Kraus vectors, $d_k$, is defined by the maximum number of Kraus operators and the size of the Hilbert space as follows:
\begin{eqnarray}
d_k&=&2d \ \times d^2 = 2d^3, \nonumber \\
d_k&=&2^{3n+1}.
\end{eqnarray}
By limiting the number of Kraus operators to one, for any qubit system, we can model the unitary channels. For any system, reducing the number of Kraus operators reduces the dimensionality of Kraus vectors. For example, general single-qubit channels result in $d_k=16$, whereas unitary single-qubit channels have $d_k=4$.

\section{\label{sec:generators_transformations}Generators and Finite Transformations for Multi-Qubit Kraus Vectors}

A key feature of this parameterization is that it provides a geometric representation of quantum channels. The set of $d$ Kraus vectors forms an \emph{orthogonal-symplectic frame}, which we call a Kraus frame $f$:
\begin{eqnarray}
    f = \{\mathbf{v}_i\}_{i=1}^d.
\end{eqnarray}
By changing the orientation of $f$, we can span the full space of Kraus operators. The transformations $M$, that transform $f$, such that
\begin{eqnarray}
    f' = M f = \{ M \mathbf{v}_i \}_{i=1}^d,
\end{eqnarray}
must preserve the orthogonal-symplectic structure so that the elemnets of $f'$ satisfy Eq.~(\ref{eq:ncomprel}), and
$f'$ is also a valid Kraus frame. Here, $M$ is a linear transformation acting on each Kraus vector $\mathbf{v}_i$, and the set of all such transformations forms a group that preserves the orthogonal-symplectic structure of the frame. 
\subsection{\label{sec:generators}Lie algebra for the Required Group}
The transformations which preserve the dot products are the orthogonal transformations, likewise to preserve the symplectic inner product we need transformations which preserve the symplectic bilinear form $S$. We can make it concrete by saying that the required transformations $M$ must satisfy Eq. (\ref{eq:ncomprel}), so the vectors in $f$ and $f'$ must follow
\begin{eqnarray}
    \mathbf{v}'_i \cdot \mathbf{v}'_j + i (\mathbf{v}'_i \cdot S \mathbf{v}'_j)=\mathbf{v}_i \cdot \mathbf{v}_j + i (\mathbf{v}_i \cdot S \mathbf{v}_j),
    \label{eq:preservation}
\end{eqnarray}
where $\mathbf{v'_i}=M\mathbf{v_i}$. Now, simplifying further, we get
\begin{eqnarray}
    {\mathbf{v}'_i}^{\top}(I + iS) \mathbf{v}'_j={\mathbf{v}_i}^{\top}(I + iS) \mathbf{v}_j,
    \label{eq:symtery} \nonumber\\
    {\mathbf{v}_i}^{\top}M^{\top}(I + iS)M \mathbf{v}_j={\mathbf{v}_i}^{\top} (I + iS)\mathbf{v}_j, \nonumber\\
    M^{\top}(I + iS)M=I + iS, \nonumber \\
    M^{\top}M + iM^{\top}SM=I + iS,
\end{eqnarray}
From this we get our required symmetry conditions
\begin{eqnarray}
    M^{\top}M=I,
    \label{eq:orthogonalS}
\end{eqnarray}
which gives us the generators of rotations with the algebra $J^{\top}=-J$, and the second
\begin{eqnarray}
    M^{\top}SM=S.
    \label{eq:symplecticS}
\end{eqnarray}
Thus, $M$ are the transformations at the intersection of rotations and symplectic group. Generally, we can write $M$ parameterized by $\theta$ as
\begin{eqnarray}
    M(\theta) = I + \theta J + O(\theta^2),
    \label{eq:tranG}
\end{eqnarray}
where $J$ is the generator of the transformation and $O$ contains all the higher order terms of $\theta$. We are interested in finding out the generators for the case where $\theta$ is very small and Eq. (\ref{eq:tranG}) becomes 
\begin{eqnarray}
    M^\top S M 
= (I + \theta J)^{\top} S \, (I + \theta J).
\label{eq:seq}
\end{eqnarray}
Substituting $M^{\top}=M^{-1}$ from Eq. (\ref{eq:orthogonalS}) into Eq. (\ref{eq:seq}), we get 
\begin{eqnarray}
    S = (I + \theta J)^{-1} S (I + \theta J), \nonumber\\
    S
= (I - \theta J)S(I + \theta J), \nonumber\\
   S J - J S = 0.
\end{eqnarray}
One last thing to note is that $S$ has the exact form of the real representation of the generator of the global phase. By requiring that $J$ are orthogonal to $S$, we can eliminate the global phase transformations from our generators. The orthogonality condition is epressed as
\begin{eqnarray}
    \operatorname{Tr}(S^{\top}J)=0.
\end{eqnarray}

\noindent We can summarize the required algebra as
\begin{eqnarray}
    J^{\top}=-J, \nonumber\\
    \left[S,J\right]=0, \nonumber\\
    \operatorname{Tr}(S^{\top}J)=0.
    \label{eq:liealgenra}
\end{eqnarray}

\subsection{\label{sec:transformations}Exponentiation of Generators for Finite Transformations}
To find the finite transformation matrices, we exponentiate the generators 
\begin{eqnarray}
    M(\theta)=e^{\theta J}.
\end{eqnarray}
By expanding the power series we get
\begin{eqnarray}
    M(\theta)=I + \theta J + \frac{\theta^2}{2!} J^2 + \frac{\theta^3}{3!} J^3 + \frac{\theta^4}{4!} J^4 + \cdots.
    \label{eq:expo}
\end{eqnarray}
Using linear algebraic methods, we computationally determined the generators satisfying the algebra in Eq. (\ref{eq:liealgenra}). We found that for any qubit system, each generator exhibits the following power structure:
\begin{eqnarray}
    J^0=I, \quad
    J^2=-P_J, \quad
    J^4=P_J, \cdots, \nonumber\\
    J^1=J, \quad
    J^3=-J, \quad
    J^5=J,\cdots.
\end{eqnarray}
$P_J$ are the projector matrices of $J$. In this way Eq. (\ref{eq:expo}) becomes
\begin{eqnarray}
    M=I + P_J \left( 1 - \frac{\theta^2}{2!} +  \cdots \right)
+ J \left( \theta - \frac{\theta^3}{3!} +   \cdots \right).
\end{eqnarray}
The final form for the transformation matrix becomes
\begin{eqnarray}
    M(\theta)=I+(\cos(\theta)-1)P_J+\sin(\theta)J.
\end{eqnarray}
This gives us a direct approach to span the space of CPTP maps by applying all the transformations with desired angles on a randomly generated Kraus frame or on an identity channel Kraus frame $K_I$ as
\begin{eqnarray}
    \label{eq:frame_prime}
    K(\theta_1,\theta_2,\cdots,\theta_n) = M(\theta_1,\theta_2,\cdots,\theta_n)K_I,
\end{eqnarray}
and a general quantum channel can be represented as
\begin{eqnarray}
    \label{eq:pchannel}\mathcal{E}_{\theta}=\mathcal{M}(M_{\theta}K_I),
\end{eqnarray}
where $\mathcal{M}$ is the map that takes a Kraus frame to a set of Kraus operators as described in section \ref{sec:parameterization}.

\section{\label{sec:methodology}Optimization of the Quasi Inverse}
\begin{figure}[h]
\centering
\includegraphics[width=1\linewidth]{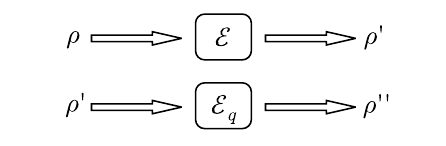}
  \caption{Original states $\rho$ are transformed by $\mathcal{E}$ into $\rho'$ and then fed into $\mathcal{E}_q$ to produce $\rho''$ which are the recovered states.}
  \label{fig:quasi_inverse_schematic}
\end{figure}

In general, quantum channels are not invertible, and only unitary channels possess exact inverses. A quasi inverse provides an approximate reversal of a noise process, undoing its effect to the greatest possible extent. Consider a general quantum channel $\mathcal{E}$ acting on a state $\rho$, producing $\rho' = \mathcal{E}(\rho)$. This output is then passed through a quasi inverse channel $\mathcal{E}_q \approx \mathcal{E}^{-1}$, yielding $\rho'' = \mathcal{E}_q(\rho')$, where $\mathcal{E}_q$ is optimized to maximize the fidelity $F(\rho, \rho'')$. The procedure is illustrated schematically in Fig. \ref{fig:quasi_inverse_schematic}.

\begin{figure}[h]
\centering
\includegraphics[width=1\linewidth]{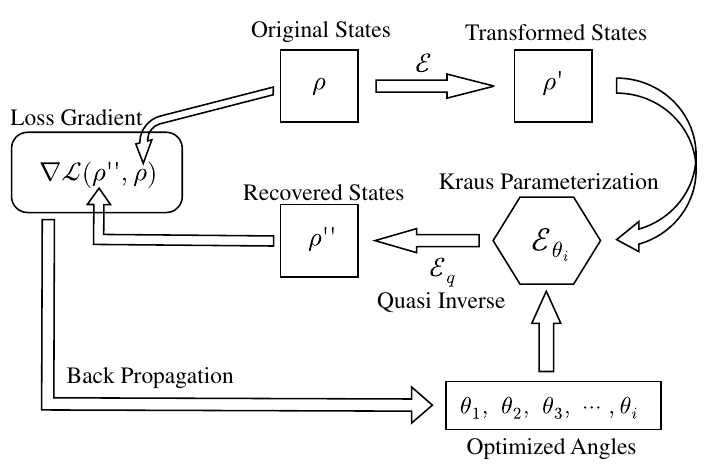}
  \caption{The process starts from original states $\rho$ which are transformed by $\mathcal{E}$ into $\rho'$ and then fed into Kraus parameterization $\mathcal{E}_{\theta_i}$ to recover to $\rho''$. The loss function is computed using $\rho$ and $\rho''$, and its gradient gives the optimized angles which are backpropagated for further optimization, finally giving the quasi inverse channel $\mathcal{E}_q$.}
  \label{fig:optimization_schematic}
\end{figure}

We prepare a sample of randomly generated mixed density matrices $\rho_i$. Then we pass them from $\mathcal{E}$, for which we want to find the quasi inverse and get \(\rho_i'\). By inserting $\rho_i'$ into the parameterized channel \(\mathcal{E}_{\theta}\) and optimizing $\theta_i$ to get the recovered states $\rho_i''=\mathcal{E}_{\theta}(\rho_i')$ such that:
\begin{eqnarray}
\mathcal{E}_q \;=\; \mathcal{E}_{\theta^\ast},
\qquad 
\theta^\ast \;=\; \arg_{\theta}(\max\, F\!\left( \mathcal{E}_{\theta}(\rho'),\, \rho \right)).
\end{eqnarray}
We perform this optimization task, which maximizes the fidelity, using a gradient decent approach with an average fidelity based loss function. A brief schematic of this approach is given in Fig. \ref{fig:optimization_schematic}.

\subsection{Loss Function and Fidelity Measure}
To quantify the performance of the predicted $\mathcal{E}_q$, we use a fidelity-based loss function defined as:
\begin{eqnarray}
\mathcal{L} = 1 - \bar{F},
\label{eq:pko23}
\end{eqnarray}
where $\bar{F}$ is the average fidelity defined as:
\begin{eqnarray}
\bar{F} = \frac{1}{N} \sum_{i=1}^N F(\rho_i'', \rho_i),
\label{eq:pko24}
\end{eqnarray}
and \( F(\rho'', \rho) \) denotes the Uhlmann fidelity \cite{nielsen}:
\begin{eqnarray}
F(\rho'', \rho) = \left( \mathrm{Tr} \sqrt{ \sqrt{\rho''} \rho \sqrt{\rho''} } \right)^2.
\label{eq:pko25}
\end{eqnarray}
Minimizing this loss function, in turn, maximizes the average fidelity. 

\subsection{Gradient Based Optimization on Kraus Sphere}
The recovered states $\rho_i''$ are functions of $\mathcal{E}_q$ and $\rho_i$. $\mathcal{E}_q$ lives on the Kraus sphere and are parameterized by a Kraus frame. In this way, the loss function gets modified to
\begin{eqnarray}
\mathcal{L}(\theta_j) = 1 - \frac{1}{N} \sum_{i=1}^N F(\mathcal{E}_{\theta_j}(\rho'_i), \rho_i).
\label{eq:pko41}
\end{eqnarray}
We use a central difference method \cite{press} to calculate the gradient with respect to each angle as:
\begin{eqnarray}
\frac{\partial \mathcal{L}}{\partial \theta_i} = \frac{\mathcal{L}(\theta_i + \varepsilon) - \mathcal{L}(\theta_i - \varepsilon)}{2\varepsilon},
\label{eq:pko26}
\end{eqnarray}
where a small value of \( \varepsilon \sim 10^{-6} \) ensures a stable numerical estimate. After computing the gradient vector, it is updated via
\begin{eqnarray}
\theta^{(t+1)} = \theta^{(t)} - \eta_0 \cdot \nabla \mathcal{L},
\label{eq:pko27}
\end{eqnarray}
where, $\eta_0$ is the base learning rate parameter. With optimized $\theta_i$ at hand and, by using Eq. (\ref{eq:pchannel}), the final output is a set of CPTP Kraus operators, defining a quasi inverse \( \mathcal{E}_q \), which maximizes the fidelity between the recovered states and the original states. 

By this method, the inverse of a wide class of quantum channels can be approximated. 

\section{\label{sec:results}Results and Discussion}
We evaluated the performance of the proposed method of approximating the quasi inverse by testing it on a range of quantum channels.

\subsection{Single-Qubit System}\label{sec:flip_channels}
For this system, we took $1000$ mixed states randomly sampled from a bloch ball. They were transformed by the noise channel and then average fidelity was measured after revovery.

Bit flip (BF) channel is the quantum analogue of a classical Not gate. It flips state $|0\rangle$ to $|1\rangle$ and vice versa with a probability of $p$. In the Kraus operator form it is written as:
\begin{eqnarray}
\nonumber
K_1 &=& \sqrt{1-p} I, \\
K_2 &=& \sqrt{p} X.
\end{eqnarray}
Likewise, the phase flip (PF) channel is a channel that flips the phase of states with some probability $p$ and thus affects the quantum interference of a qubit. It is described by the following Kraus operators:
\begin{eqnarray}
\nonumber
K_1 &=& \sqrt{1-p} I, \\
K_2 &=& \sqrt{p} Z.
\end{eqnarray}
The mixture of these gates, which is called a bit-phase flip (BPF) channel, induces both noises with equal probality $p$ in a state and is described as:
\begin{eqnarray}
\nonumber
K_1 &=& \sqrt{1-p} I, \\
K_2 &=& \sqrt{p} Y,
\end{eqnarray}
where $I,X,Y,Z$ are the pauli basis.

\begin{figure}[h]
    \centering
    \includegraphics[width=1\linewidth]{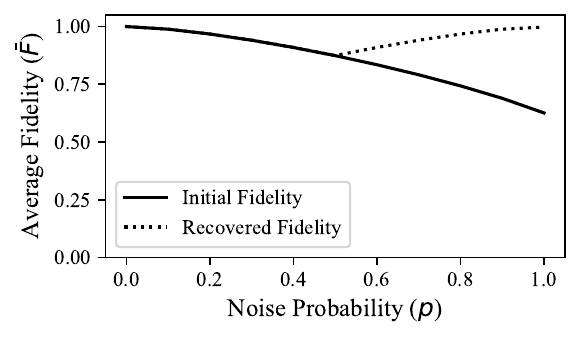}
    \caption{Fidelity recovery curve for the single-qubit bit flip channel generated with varying noise strength $p$. High recovery occurs for strong noise after $p=0.5$.}
    \label{fig:fig3a_bit_flip_fidelity_plot}
\end{figure}

For all three classical error channels with noise strength $p = 0.8$, this quasi inverse method demonstrated substantial fidelity recovery. Bit flip improved from 0.6989 to 0.9621 with the following Kraus operator representation for its quasi invers:
\begin{eqnarray}
\label{eq:qibf}
K_1 &=&
\left[
\begin{array}{cc}
0.0149 - 0.0027\,\mathrm{i} & -0.9971 + 0.0744\,\mathrm{i} \\
-0.9966 + 0.0807\,\mathrm{i} & -0.0151 - 0.0004\,\mathrm{i}
\end{array}
\right], \nonumber\\
K_2 &=& K_3 = K_4 =
\left[
\begin{array}{cc}
0 & 0 \\
0 & 0
\end{array}
\right].
\end{eqnarray}
The fidelity recovery trend with respect to the noise strength can be seen in Fig. \ref{fig:fig3a_bit_flip_fidelity_plot}.

\begin{figure}[h]
    \centering
    \includegraphics[width=1\linewidth]{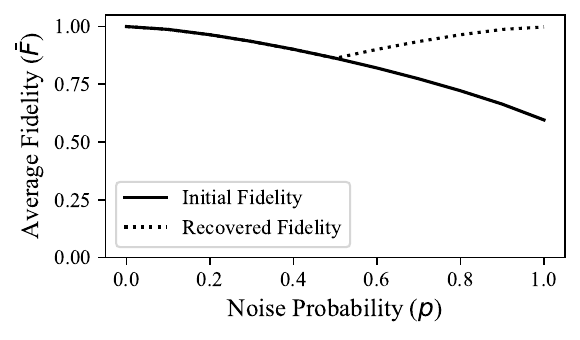}
    \caption{Fidelity recovery curve for the single-qubit phase flip channel generated with varying noise strength $p$.}
    \label{fig:fig3b_phase_flip_fidelity_plot}
\end{figure}

Phase flip channel recovered from 0.7051 to 0.9642 and has the following Kraus operators
\begin{eqnarray}
\label{eq:qipf}
K_1 &=&
\left[
\begin{array}{cc}
-0.0226 - 0.9996\,\mathrm{i} & -0.0152 - 0.0072\,\mathrm{i} \\
\;\;\;0.0149 - 0.0079\,\mathrm{i} & \;\;\;0.0225 + 0.9996\,\mathrm{i}
\end{array}
\right], \nonumber\\
K_2 &=& K_3 = K_4 =
\left[
\begin{array}{cc}
0 & 0 \\
0 & 0
\end{array}
\right].
\end{eqnarray}
Fidelity recovery trend for phase flip channel is given in Fig. \ref{fig:fig3b_phase_flip_fidelity_plot}

\begin{figure}[h]
    \centering
    \includegraphics[width=1\linewidth]{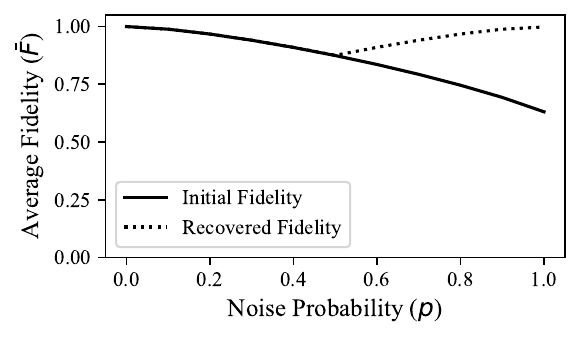}
    \caption{Fidelity recovery curve for the single-qubit bit-phase flip channel generated with varying noise strength $p$.}
    \label{fig:fig3c_bit_phase_flip_fidelity_plot}
\end{figure}

Lastly the bit-phase flip channel had a recovery from 0.7259 to 0.9655. The Kraus operators for the quasi inverse of this channel are:
\begin{eqnarray}
\label{eq:qibpf}
K_1 &=&
\left[
\begin{array}{cc}
-0.0012 - 0.0011\,\mathrm{i} & \;\;0.3990 + 0.9170\,\mathrm{i} \\
-0.4005 - 0.9163\,\mathrm{i} & \;\;0 - 0.0016\,\mathrm{i}
\end{array}
\right], \nonumber\\
K_2 &=& K_3 = K_4 =
\left[
\begin{array}{cc}
0 & 0 \\
0 & 0
\end{array}
\right].
\end{eqnarray}
Recovered fidelity curve for bit-phase flip channel is shown in Fig. \ref{fig:fig3c_bit_phase_flip_fidelity_plot}.

For higher noise probabilities, all of these channels show good recovery using the quasi inverse. Moreover, we use parameterization of general channels to optimize the quasi inverse channels which all come out to be unitary,  as seen in Eq. (\ref{eq:qibf}), (\ref{eq:qipf}) and (\ref{eq:qibpf}). These results agree with the findings of \cite{karimipour2020} and \cite{faizan2025}.

\subsection{Two-Qubits System}
We took $100$ mixed states using Bures random sampling over Bloch ball for the training data. We find quasi inverse for two-qubit bit flip, phase flip and bit-phase flip channels as well. Every qubit can undergo a flip independently, so to model the collective action of any of the flip channels on each qubit individually, a tensor product structure is used. Each qubit has its own set of Kraus operators as seen in section (\ref{sec:flip_channels}), and for an $n$-qubit system, we can construct the full set of Kraus operators by taking all possible combinations of the single-qubit Kraus operators and forming their tensor products as \cite{gottesman}:
\begin{eqnarray}
    P = K^{(1)} \otimes K^{(2)} \otimes \cdots \otimes K^{(n)},
\end{eqnarray}
where $K^{(i)}$ can be any of the single-qubit Kraus operators for the $i^{\mathrm{th}}$ qubit. This quantum channel is then applied as  
\begin{eqnarray}
    \mathcal{E}(\rho)=p\sum_P P \rho P^\dagger,
\end{eqnarray}
where the sum is taken over the Kraus operators $P$ of the $n$-qubit system. At noise strength \( p = 0.8 \), we computed quasi inverse of these two-qubit channels which were all unitary channels as well. 

\begin{figure}[h]
    \centering
\includegraphics[width=1\linewidth]{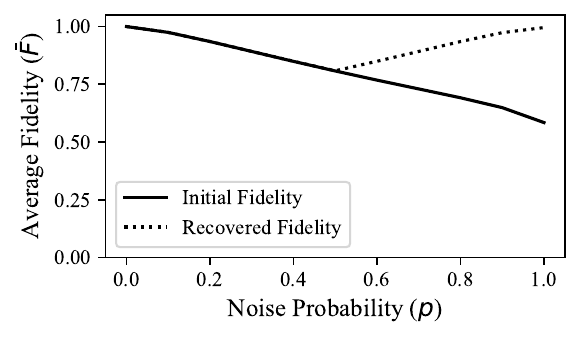}
    \caption{Fidelity recovery curve for the two-qubit bit flip channel against the noise probability $p$.}
    \label{fig:fig5a_two_qubit_bit_flip_fidelity_plot}
\end{figure}

Bit flip improved from 0.6884 to 0.9295. The fidelity vs noise strength curve for this channel is given in Fig. \ref{fig:fig5a_two_qubit_bit_flip_fidelity_plot}

Phase flip channel recovered from 0.6738 to 0.9342 and has the following recovery trend against varrying noise strength in Fig. \ref{fig:fig5b_two_qubit_phase_flip_fidelity_plot}:
\begin{figure}[h]
    \centering
    \includegraphics[width=1\linewidth]{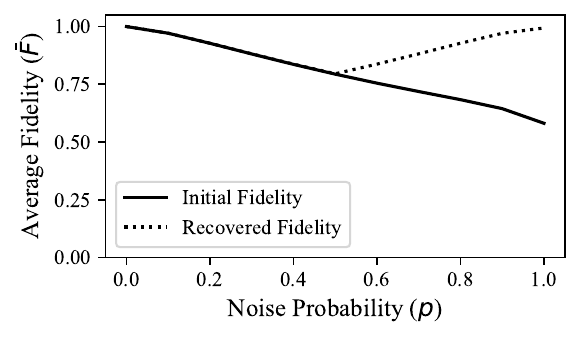}
    \caption{Fidelity recovery curve for the two-qubit phase flip channel.}
    \label{fig:fig5b_two_qubit_phase_flip_fidelity_plot}
\end{figure}

\begin{figure}[h ]
    \centering
    \includegraphics[width=1\linewidth]{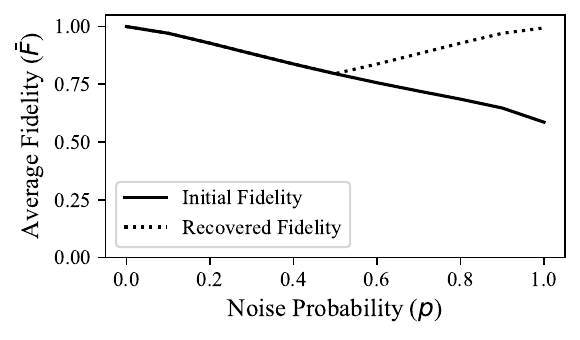}
    \caption{Fidelity recovery curve for the two-qubit bit-phase flip channel.}
    \label{fig:fig5c_two_qubit_bit_phase_flip_fidelity_plot}
\end{figure}

Lastly, the bit-phase flip channel had a recovery from 0.6637 to 0.9655. The fidelity curve is given in Fig. \ref{fig:fig5c_two_qubit_bit_phase_flip_fidelity_plot}.

We clearly observe similar recovery trends for one and two-qubit systems. We also see that for both systems, even after using the general channel parameterization scheme, we essentially get unitary channels for quasi inverses. Finally, we also optimized the quasi inverse for depolarizing channels. The result was identity channel, and no recovery was observed.

\section{\label{sec:conclusions}Concluding Remarks}
In this work, we introduced a differentiable geometric parameterization of Kraus operators for arbitrary multi-qubit systems. By mapping each Kraus operator to a set of orthogonal–symplectic Kraus vectors on a hypersphere, the framework ensures complete positivity and trace preservation by construction. This geometric structure, together with the derived Lie algebra and finite orthogonal–symplectic transformations, supplies a compact and physically motivated method to traverse the full CPTP manifold. Using this machinery, we formulated the quasi-inverse problem as a fidelity-maximization task and optimized over the Kraus sphere using gradient-based techniques to obtain approximate inverse channels.

The methodology offers clear conceptual and computational advantages. The hyperspherical Kraus representation is naturally differentiable, which is essential for machine-learning–based channel learning tasks. The Lie-group interpretation of CPTP-preserving transformations further exposes an elegant algebraic structure, enabling systematic exploration of the channel space. This makes the framework well-suited for variational algorithms, Monte-Carlo sampling, and noise-learning protocols where both physical validity and smoothness are necessary.

Our results demonstrate that the proposed approach reliably learns high-fidelity quasi-inverse channels for single-qubit and two-qubit quantum channels. Across bit flip, phase flip, and bit-phase flip channels, the learned quasi-inverses consistently recovered the original states with high fidelity. A noteworthy and theoretically consistent outcome is that optimization over the full general CPTP parameterization still converges to unitary quasi-inverse channels, matching known analytical expectations. For channels such as the depolarizing channel, where no quasi-inverse exists, the method correctly identifies the identity channel, further validating the robustness of the approach.

While this approach has no redundant degrees of freedom when parameterizing unitary channels, the general CPTP parameterization does contain redundancies due to the inherent gauge freedom in Kraus representations: different sets of Kraus operators can describe the same physical channel. The Lie algebra we derived naturally includes generators corresponding to unitary mixing among Kraus operators, introducing additional degrees of freedom for non-unitary channels. A careful analysis of how Kraus vectors transform under such unitary mixing, combined with a suitable gauge-fixing procedure, could eliminate redundancy and significantly improve computational efficiency.

Finally, a promising future direction is to leverage this geometric framework to characterize the gauge freedom in Kraus representations, which leaves the physical channel invariant, suggesting the existence of conserved geometric structures under such transformations. Identifying such geometric invariants would allow one to determine whether two seemingly different Kraus representations correspond to the same CPTP channel. Developing this geometric characterization would not only deepen our conceptual understanding of quantum channels but also provide a practical tool for channel classification, benchmarking, and machine-learning applications where distinguishing physically equivalent channels is essential. 

Additionally, exploring whether continuous trajectories on the Kraus sphere correspond to dynamical evolution of open quantum systems or specific noise processes could offer a novel dynamical interpretation of the framework.

 \section*{Acknowledgements}
The authors acknowledge discussions with Sameen Aziz on the evaluation of quasi inverse using neural networks.

\end{document}